\documentclass[aps,prb,twocolumn,superscriptaddress,amsmath,amssymb,showpacs]{revtex4-1}
\usepackage{amssymb}
\usepackage{graphicx}
\usepackage{bm}
\usepackage{epsfig}
\usepackage{ulem}
\usepackage{color}
\usepackage{amsmath}
\usepackage{ dsfont }
\usepackage{float}
\usepackage[table]{xcolor}
\usepackage{xcolor,colortbl}
\usepackage{tabularx}
\usepackage{ulem}
\usepackage{mathtools}

\newcommand{\SRO}{Sr$_2$RuO$_4\:$}

\def\bk{{\bold{k}}}
\def\up{{\uparrow}}
\def\down{{\downarrow}}
\def\dg{{^\dagger}}
\def\m1{{^{-1}}}

\begin{document}
\normalem

%%%%%%%%%%%%%%%%%%%%%%%%%%%%%%%%%%%%%%%%%

\title{Identifying detrimental effects for multi-orbital superconductivity - Application to Sr$_2$RuO$_4$}

\author{Aline Ramires}
\affiliation{Institute for Theoretical Studies, ETH Z\"{u}rich, CH-8092, Z\"{u}rich, Switzerland.}

\author{Manfred Sigrist}
\affiliation{Institute for Theoretical Physics, ETH Zürich, CH-8093, Zürich, Switzerland}

\date{\today}

\begin{abstract}
We propose a general scheme to probe the compatibility of arbitrary pairing states with a given normal state Hamiltonian by the introduction of a concept called \emph{superconducting fitness}. This new quantity gives a direct measure of the suppression of the superconducting critical temperature in presence of key symmetry breaking fields. A merit of the \emph{superconducting fitness} is that it can be used as a tool to identify non-trivial mechanisms to suppress superconductivity under various external influences, in particular, magnetic fields or distortions, even in complex multi-orbital systems. In the light of this new concept we analyse the multi-band superconductor Sr$_2$RuO$_4$ and propose a new mechanism for the suppression of superconductivity in multi-orbital systems, which we call \emph{inter-orbital effect}, as a possible explanation for the unusual limiting feature observed in the upper critical field of this material.
\end{abstract}

\maketitle

%%%%%%%%%%%%%%%%%%%%%%%%%%%%%%%%%%%%%%%%%

\section{Introduction}

%Multi-orbital effects
% 1. Shape the SC state
%  2. New effects in presence of symmetry breaking fields
The study of superconductivity (SC) in complex multi-orbital systems has been motivated by several classes of materials, including cuprate\cite{Bed}, Fe-based\cite{Kam}, heavy fermion\cite{Ste} and ruthenade\cite{Mae} superconductors. In in these systems several orbitals contribute to the bands crossing the Fermi level, and from the microscopic Hamiltonian in the orbital basis it is usually not immediately clear how inter-orbital effects, such as spin-orbit coupling (SOC), help shape the character of the superconducting state.  Also, the effects of external symmetry breaking fields on the stability of different superconducting states in presence of many orbitals is not generally understood. Spin polarization induced by an external magnetic field and anti-symmetric spin-orbit coupling  originated from the absence of inversion symmetry are two examples of effects that are detrimental to Cooper pairing in the spin singlet and spin triplet channels, respectively\cite{And1,And2,Sig}. The generalization of this knowledge to multi-orbital systems is often not straightforward, and this work presents an attempt to address this point with a simple concept which we call \emph{superconducting fitness}.

%Brief intro to criterium
Here we propose a modified commutator as a measure of the incompatibility between an arbitrary pairing state described by the gap matrix $\hat{\Delta}(\bk)$ and a given normal state Hamiltonian $H_0(\bk)$:
\begin{eqnarray}\label{Crit}
\hspace{-0.25cm}[H_0(\bk) ,\hat{\Delta}(\bk) ]^*=F(\bk)  (i\sigma_2),
\end{eqnarray}
where we define $F(\bk)$, the \emph{superconducting fitness}, which quantifies the incompatibility between the gap and the underlying band structure. Note that the use of a commutator assumes we write the Hamiltonian and the gap function in matrix form, acting on \emph{multi-orbital Nambu spinors}. A complete motivation and definition of the quantities above is given in section \ref{GenCri}. A merit of this concept is that it can be easily applied to the stability analysis of superconducting states in multi-band systems, to establish gap structures favourable within a given complex band structure. Also, it can be used as a tool to identify non-trivial mechanisms to suppress superconductivity under various external influences, in particular, magnetic fields or distortions. This concept does, however, not address directly any aspects concerned with the pairing mechanism.

Recently, Fischer \cite{Fis} revisited a similar analysis and applied it to the study of Fe-based SC. The idea is ultimately based on the BCS theory, which requires pairing of electrons with same energy and total momentum equals to zero in order to develop an instability towards a superconducting state for an arbitrarily small attractive interaction. Here we formally develop a generalized version of this idea which correctly addresses triplet superconductivity, obtaining a simple closed form for the effects of symmetry breaking fields in the critical temperature, and show how it can be generalized to multi-orbital systems.

% Paper organization
This paper is organized as follows: In section \ref{GenCri} we show how Fischer's criterium \cite{Fis} is generalized to Eq.~\ref{Crit} and can now be applied to triplet SC and to systems in presence of SOC and transverse magnetic fields. In section \ref{Val} we show how this concept is in accordance with weak coupling results by evaluating the effect of the presence of key symmetry breaking fields on the superconducting critical temperature $T_c$ for a single band SC using the Green's function (GF) formalism. Next, in section \ref{MultiBand} we generalize the discussion to multi-band systems.
Amongst the the multi-orbital superconductors which are still warmly debated today is \SRO\!\cite{Mac}. In section \ref{SRO} we discuss what the \emph{superconducting fitness} suggests as the most compatible SC state and its stability under symmetry breaking fields. In particular, we propose a new mechanism for the suppression of the superconducting state in multi-orbital systems and discuss how it can be related to the unusual limiting feature observed in the upper critical field in \SRO\!.

%%%%%%%%%%%%%%%%%%%%
\section{The concept of Superconducting Fitness}\label{GenCri}

In multi-orbital superconducting systems, the effective mean field Hamiltonian can be written as:
\begin{eqnarray}\label{MFH}
H_{MF}&=&\sum_\bk \Psi_\bk^\dagger \begin{pmatrix}
 H_{0}(\bk)& \Delta(\bk)\\
\Delta^\dagger(\bk) & - H^*_{0} (-\bk)
\end{pmatrix}
\Psi_\bk,
\end{eqnarray}
in terms of the \emph{multi-orbital Nambu spinors}:
\begin{eqnarray}
\Psi_\bk^\dagger = (\psi^\dagger_{\bk}, \psi_{-\bk}^T) ,
\end{eqnarray}
with
\begin{eqnarray}\label{Basis}
\psi^\dagger_{\bk}= (a_{1\bk \up}^\dagger, \hspace{0.1cm} a_{1\bk \down}^\dagger ,..., a_{n\bk \up}^\dagger, \hspace{0.1cm} a_{n\bk \down}^\dagger).
\end{eqnarray}
Here $a_{m\bk\sigma}^\dagger$ and $a_{m\bk\sigma}$ creates and annihilates an electron in orbital $m$, with momentum $\bk$ and spin $\sigma=\{\uparrow,\downarrow\}$, respectively. Here $H_0(\bk)$ and $ \Delta(\bk)$ are $2n \times 2n$ matrixes where $n$ is the number of orbitals. $H_0(\bk)$ is the bare Hamiltonian in the orbital basis which is not necessarily diagonal and can include inter-orbital hopping and spin-orbit coupling terms; $ \Delta(\bk)$ is the gap function which can describe both singlet and triplet pairing.

In general, there is a unitary transformation $U_\bk$ which diagonalizes $H_0(\bk)$:
\begin{eqnarray}
\psi^\dagger_{\bk} H_{0}(\bk) \psi_{\bk}  \xrightarrow{U_\bk}  (\psi^B_{\bk})^\dagger H^B_0(\bk)\psi^B_{\bk},
\end{eqnarray}
where $\psi^B_{\bk}=U _\bk \psi_{\bk} $ is the rotated band basis in which $H^B_0(\bk)=U_\bk H_0(\bk) U_\bk^\dagger$ is diagonal. The gap matrix, by connecting particle-particle spaces, transforms differently:
\begin{eqnarray}
\psi^\dagger_{\bk} \Delta(\bk) \psi_{-\bk}^*  \xrightarrow{U_\bk}  (\psi^B_{\bk})^\dagger \Delta^B(\bk)(\psi^B_{-\bk})^*,
\end{eqnarray}
where $\Delta^B(\bk) =U_\bk \Delta(\bk) U^T_{-\bk}$ is not necessarily diagonal.

The development of superconductivity is usually guaranteed by the presence key symmetries: time-reversal symmetry for singlet states and inversion symmetry for triplet states \cite{And1,And2}. Once these symmetries are broken the degeneracy of the states of the electrons to be paired is also lost. Pairing now happens between electrons at different energies, or different bands, and the SC state is not as stable since now one needs a finite attractive interaction to overcome this energy difference. With the general idea that inter-band pairing usually does not lead to the most stable superconducting state (be it originated from symmetry breaking or not), now we analyze what conditions the assumption of pure intra-band pairing gives us for the Hamiltonian and gap function in the orbital basis.

In order to have only intra-band pairing $\Delta^B(\bk)$ should be block diagonal with $n$ $2 \times 2 $ blocks. To get some intuition on the origin of the concept of \emph{superconducting fitness}, we consider the minimal multi-band problem consisting of two bands. In presence of time reversal and inversion symmetries $H_0(\bk)$ has doubly degenerate states with energy $\epsilon_a$, where $a$ is the band label, and therefore has a structure with $2 \times 2$ blocks proportional to the identity $\sigma_0$. Note that for an arbitrary gap matrix with intra- $\Delta_{a}$ and inter- $\Delta_{ab}$ band components, we have, omitting the momentum dependence:
\begin{eqnarray}\label{MFH2}
H^B_0 =\begin{pmatrix}
\epsilon_1 \sigma_0 & 0\\
0& \epsilon_2 \sigma_0
\end{pmatrix},\hspace{0.5cm}
\Delta^B  &=& \begin{pmatrix}
\Delta_1 & \Delta_{12}\\
\Delta_{21}& \Delta_2
\end{pmatrix}.
\end{eqnarray}
Note that these matrices do not commute for finite inter-band pairing, unless the artificial condition $\epsilon_1=\epsilon_2$ is satisfied. On the other hand, in case $\Delta_{ab}=0$, $\Delta^B$ is block diagonal and commutes with the diagonalized bare Hamiltonian $H^B_0$ in the band basis.

We can now look at the condition for absence of inter-band pairing from the orbital basis perspective. Restoring the momentum dependence, using the unitary transformation introduced above and the fact that $H^B_0(\bk)$ and $\Delta^{B}(\bk)$ commute in case of pure intra-band pairing, we can write:
\begin{eqnarray}
H_0(\bk) \Delta(\bk) &=& U^\dagger_\bk H^B_0(\bk) U_{\bk}U^\dagger_\bk  \Delta^{B}(\bk) U_{-\bk}^* ,\\
&=&U^\dagger_\bk H^B_0(\bk)  \Delta^{B}(\bk) U_{-\bk}^*\nonumber\\
&=&U^\dagger_\bk  \Delta^{B}(\bk)  H^B_0(\bk) U_{-\bk}^*\nonumber\\
&=&U^\dagger_\bk  \Delta^{B}(\bk)U_{-\bk}^* U_{-\bk}^T   H^B_0(\bk) U_{-\bk}^*.\nonumber
\end{eqnarray}
Here we can identify the first three factors in the last line with $\Delta(\bk)$. For the last three factors, we use inversion symmetry (already assumed above to guarantee the double degeneracy of the bands) and the fact that the eigenvalues of $H_0^B(\bk)$ are real to show:
\begin{eqnarray}
H^*_0(-\bk) &=& (U^\dagger_{-\bk} H^B_0(-\bk) U_{-\bk})^* \\
&=& U^T_{-\bk} H^B_0(-\bk) U_{-\bk}^* \nonumber\\
&=&  U^T_{-\bk} H^B_0(\bk) U_{-\bk}^*,\nonumber
\end{eqnarray}
so we can write:
\begin{eqnarray}
H_0(\bk) \Delta(\bk) - \Delta(\bk)H^*_0(-\bk) = 0.
\end{eqnarray}
If $H_0(\bk)$ and $\Delta(\bk)$ satisfy this condition, the system develops only intra-band pairing and consequently has a more stable SC phase. We understand this equality as a probe of the compatibility of arbitrary pairing states with a given normal state Hamiltonian, and propose the short notation of Eq.~\ref{Crit} to refer to it from now on:
\begin{eqnarray}\label{CritDef}
\hspace{-0.25cm}[H_0(\bk) ,\hat{\Delta}(\bk) ]^*&=&H_0(\bk) \hat{\Delta}(\bk) - \hat{\Delta}(\bk)H^*_0(-\bk)\\\nonumber&=&F(\bk)(i\sigma_2),
\end{eqnarray}
where we introduced the matrix $F(\bk)$, the \emph{superconducting fitness}, since in general situations this quantity is not equal to zero and gives a direct measure of the suppression of $T_c$, as will be shown in the next section. Here we write the \emph{superconducting fitness} in terms of the hatted gap matrix $\hat{\Delta}(\bk)$, which factors out the magnitude of the complex order parameter and will be defined in the next section. We also factor out $(i\sigma_2)$ for convenience in the following discussion. Note that the \emph{superconducting firtness} falls back into Fischer's condition\cite{Fis}, $[H_0(\bk), \Delta(\bk)]=0$, only in presence of inversion symmetry, $H_0(-\bk)=H_0(\bk)$, and for a real Hamiltonian $H^*_0(\bk)= H_0(\bk)$ (what does not apply in presence of SOC, transverse magnetic fields and inversion symmetry breaking).

%%%%%%%%%%%%%%%%%%%%%%%%%%%%%%%%%%%%%%%%%
\section{Superconducting Fitness in the single-orbital scenario}\label{Val}

In this section we discuss the concept of \emph{superconducting fitness} in the simple single-band scenario.  The non-interacting Hamiltonian for a single band system can be written as $H_0(\bk)=\epsilon(\bk) I_2$, in the basis $\psi^\dagger_{\bk}= (a_{\bk \up}^\dagger, \hspace{0.1cm} a_{\bk \down}^\dagger)$, and can be perturbed by symmetry breaking fields introduced as 
\begin{eqnarray}\label{deltaH}
\delta H(\bk)=\mathbf{s}(\bk)\cdot \boldsymbol{\sigma}_T,
\end{eqnarray}
where $\boldsymbol{\sigma}_T=(\sigma_1,\sigma_2,\sigma_3)$ is a vector formed by the three Pauli matrices.  Time reversal symmetry breaking can be implemented as an external magnetic field $\bold{h}$, in which case $\mathbf{s}(\bk)=-\bold{h}$, where $\bold{h}=(h_x,h_y,h_z)$; while inversion symmetry breaking can be introduced as a Rashba-type SOC $\mathbf{s}(\bk)=\mathbf{g}(\bk)$, where $\bold{g}(\bk)$ is a 3-component vector and an odd function of $\bk$.

We write the gap matrix in the general form:
\begin{eqnarray}
\Delta(\bk) &=& \sum_{i=1}^{d_\Gamma} \phi_i [\mathbf{\Psi}_i^{\Gamma}(\bk)\cdot \boldsymbol{\sigma}(i\sigma_2) ],
\end{eqnarray}
where the index $i$ labels the normalized basis functions $\mathbf{\Psi}_i^{\Gamma}(\bk)$ of the $d_\Gamma$-dimensional irreducible representation $\Gamma$ of the symmetry group of the system. Here $\phi_i$ are complex order parameters, and we assume we already know the irreducible representation which will lead to the highest critical temperature. Without loss of generality, from now on we assume the dimensionality of the irreducible representation of interest to be equal to one for a more clear notation. Here we define the hatted gap matrix $\hat{\Delta}(\bk)$, factoring out the magnitude of the order parameter from the gap matrix:
\begin{eqnarray}\label{GapM}
\Delta(\bk) &=&\phi [\mathbf{\Psi}^{\Gamma}(\bk)\cdot \boldsymbol{\sigma}(i\sigma_2) ], \\ \nonumber
&=&|\phi|  \hat{\Delta}(\bk).
\end{eqnarray}
Note that $\Delta(\bk)$ carries information about the phase of the order parameter. The basis function $\mathbf{\Psi}^{\Gamma}(\bk)$ encodes the momentum structure of the gap and is in fact a four-dimensional vector $\mathbf{\Psi}^{\Gamma}(\bk)=(d_0(\bk),d_x(\bk),d_y(\bk),d_z(\bk))$ such that it can describe both singlet or triplet SC, and $\boldsymbol{\sigma}=(\sigma_0,\sigma_1,\sigma_2,\sigma_3)$ (here $\sigma_0$ is the $2 \times 2$ identity matrix). For a singlet SC we have $d_0(\bk)=d_0(-\bk)\neq 0$ and $d_{x,y,z}(\bk)=0$, while for a triplet we have $d_0(\bk)=0$ and $d_{x,y,z}(\bk)=-d_{x,y,z}(-\bk)$, the last not all concomitantly zero.

For singlet SC the hatted gap matrix can be written as:
\begin{eqnarray}
\hat{\Delta}_S(\bk)=\hat{\phi}d_0(\bk) (i\sigma_2),
\end{eqnarray}
where $\hat{\phi}= \frac{\phi}{|\phi|}$ and we can evaluate the \emph{superconducting fitness} defined in Eq.~\ref{CritDef}:
\begin{eqnarray}
F_{S}(\bk) = (1+\tilde{p}) \hat{\phi} d_0(\bk) \delta H (\bk),
\end{eqnarray}
where $\tilde{p}$ is the parity of the perturbation ($\delta H (-\bk) = \tilde{p} \delta H (\bk)$). In case of inversion symmetry breaking $\tilde{p}=-1$ and $F_S(\bk)=0$, indicating that the absence of this symmetry is not detrimental to singlet superconductivity; whereas for magnetic field $\tilde{p}=1$, and explicitly:
\begin{eqnarray}
F_{S}(\bk) = - 2 \hat{\phi} d_0(\bk) \bold{h}\cdot\boldsymbol{\sigma}_T,
\end{eqnarray}
what indicates that magnetic field, irrespective of its direction, is detrimental to singlet superconductors due to time-reversal symmetry breaking.

In an analogous fashion we can analyse the stability of triplet SC, in which case the gap matrix can be parametrized as:
\begin{eqnarray}
\hat{\Delta}_T(\bk)= \hat{\phi} \left(\mathbf{d}_T(\bk)\cdot \boldsymbol{\sigma}_T \right)(i\sigma_2),
\end{eqnarray}
where $\mathbf{d}_T(\bk)=(d_x(\bk),d_y(\bk),d_z(\bk))$ is a 3-component complex vector and an odd function in $\bk$ and $\boldsymbol{\sigma}_T=(\sigma_1,\sigma_2,\sigma_3)$. In this case:
\begin{eqnarray}
F_{T}(\bk) &=& (1+\tilde{p})\hat{\phi} \left( \mathbf{s}(\bk) \cdot \mathbf{d}_T(\bk)\right) I_2 \\ \nonumber
&+& i (1-\tilde{p})\hat{\phi} \left( \mathbf{s}(\bk) \times \mathbf{d}_T(\bk)\right) \cdot \boldsymbol{\sigma}_T,
\end{eqnarray}
so explicitly, for time-reversal symmetry breaking:
\begin{eqnarray}\label{Trip1}
F_{T}(\bk) &=& -2\hat{\phi} \mathbf{h}\cdot \mathbf{d}_T(\bk) I_2,
\end{eqnarray}
and for inversion symmetry breaking:
\begin{eqnarray}\label{Trip2}
F_{T}(\bk) &=&2 i \hat{\phi}\left( \mathbf{g}(\bk) \times \mathbf{d}_T(\bk)\right) \cdot \boldsymbol{\sigma}_T.
\end{eqnarray}

This indicates that triplet SC is destabilized by a magnetic field with a component in the same direction as $\bold{d}_T(\bk)$ or by inversion symmetry breaking, except when the vectors $\bold{g}({\bk})$ and $\bold{d}_T(\bk)$ are parallel. 

The results above can be understood in terms of Anderson's theorems\cite{And1,And2} and weak coupling approaches\cite{Sig}. Note, though, that the \emph{superconducting fitness} provides an unified framework to study the stability of different superconducting states in presence of several external perturbations. In the the following subsection we derive a direct relation between $F(\bk)$ and the suppression of the critical temperature, which is valid for any SC state and symmetry breaking field. A generalization to multi-band systems is discussed in the following section.

%%%%%%%%%%%%%%%%%%%%
\subsection{Relation to the suppression of $T_c$}\label{TC}

In this section we derive an explicit form for the suppression of the superconducting critical temperature as a function of the quantity $F(\bk)$, introduced in Eq.~\ref{CritDef}. We start with the linearized gap equation \cite{Min}:
\begin{eqnarray}\label{LGap}
\Delta_{s_1s_2}(\bk) &=& - T \sum_{\bk',n, \{s'\}} V_{s_1s_2s_{1'}s_{2'}}(\bk,\bk') G_{s_{2'}s_{3'}}(\bk',i\omega_n)  \nonumber \\  &&\times \,\Delta_{s_{3'}s_{4'}}(\bk')G^T_{s_{4'}s_{1'}}(-\bk',-i\omega_n),
\end{eqnarray}
with a sum over the primed momentum $\bk' $, Matsubara frequencies $\omega_n=(2n+1)\pi T$ (where $T$ is the temperature and $n$ an integer) and the set of primed spin indexes $ \{s'\}=s_1', s_2', s_3', s_4' $.  We can write the interaction matrix in spectral form:
\begin{eqnarray}
V_{s_1s_2s_{1'}s_{2'}}(\bk,\bk') = v \hat{\Delta}_{s_1s_2} (\bk) \hat{\Delta}^\dagger_{s_{1'}s_{2'}}(\bk' ),
\end{eqnarray}
introducing $v$ as the coupling strenght in the SC channel of interest.
Inserting this explicit form of the interaction matrix and the general form of the gap matrix introduced in Eq.~\ref{GapM} above into the linearized gap equation, this can be simplified to:
\begin{eqnarray}\label{SimpleGap}
1=&& - Tv  \sum_{\bk,n}Tr \left[  \hat{\Delta}^\dagger(\bk) G(k)  \hat{\Delta}(\bk)G^T(-k)\right],
\end{eqnarray}
where we used the short notation $G(k)=G(\bk,i\omega_n)$ for the Green's function, which refers to the system plus any symmetry breaking perturbation. If we split the Hamiltonian as:
\begin{eqnarray}
H(\bk)=H_0(\bk)+\delta H (\bk) = \epsilon(\bk) \sigma_0 + \mathbf{s}(\bk)\cdot \boldsymbol{\sigma},
\end{eqnarray}
such that we make explicit the effects of symmetry breaking fields in $\delta H (\bk)$, we can expand $G (\bk,i\omega_n)$ in $\delta H (\bk)$:
\begin{eqnarray}
G(\bk,i\omega_n) &=&   \frac{1}{i\omega_n - H_0(\bk)-\delta H (\bk)} \\
\nonumber &=& G_0(\bk,i\omega_n)  \sum_{p=0}^\infty  \left( G_0(\bk,i\omega_n)\delta H(\bk)\right)^p,
\end{eqnarray}
where $G_0(\bk,i\omega_n)=( i\omega_n- \epsilon_\bk)^{-1} \sigma_0$ is the bare GF for the unperturbed system,  which is proportional to the identity for the single band problem.

Inserting this expansion in the gap equation, keeping only terms up to second order in $\delta H(\bk)$, and using properties of the Hamiltonian and gap matrix to simplify the traces, the linearized gap equation can be simplified further. After performing the sum over momenta and Matsubara frequencies (details can be found in Appendix \ref{AppTC}) we find:
\begin{eqnarray}\label{TC}
T_c&\sim& T_c^0\left(1 -\frac{7 \xi(3)}{64 \pi^2} \frac{ \left\langle  Tr|F(\bk )|^2  \right\rangle_{FS}}{(T_c^0)^2}\right),\nonumber\\
\end{eqnarray}
Here $T_c$ and $T_{c0}$ are the critical temperatures in presence and absence of the symmetry breaking fields, respectively, and $\langle...\rangle_{FS}$ denotes the average over the Fermi surface. This result is in agreement with the literature\cite{And1,And2,Sig}, but here we find an unified expression that accounts for the suppression of $T_c$ by different kinds of symmetry breaking fields and for an arbitrary SC state in terms of the \emph{superconducting fitness} $F(\bk)$.

%%%%%%%%%%%%%%%%%%%%%%%%%%%%%%%%%%%%%%%%%
\section{Generalization to Multi-Orbital systems}\label{MultiBand}

The generalization to multi-orbital systems follow similar lines as the single band case. In case of $n$ orbitals, both the non-interacting Hamiltonian and the gap matrix can be written using as basis matrices the $n^2-1$ generators of the $SU(n)$ group in the fundamental representation plus the identity matrix. Here we refer to these matrices as $\lambda_i$, with $i=\{0,1,...,n^2-1\}$, where $\lambda_0$ is the n-dimensional identity matrix. The non-interacting Hamiltonian can be written as:
\begin{eqnarray}
H_0(\bk) = \sum_{ij}a_{ij}(\bk)\lambda_i \otimes \sigma_j,
\end{eqnarray}
where $\sigma_{j}$ are the Pauli matrices plus the 2-dimensional identity with $j=\{0,1,2,3\}$. We introduced the parameters $a_{ij}(\bk)$ which encode all the information about hoppings and SOC. In absence of symmetry breaking fields and SOC terms, this Hamiltonian reduces to
\begin{eqnarray}
H_0(\bk) = \mathbf{a}(\bk)\cdot \boldsymbol{\lambda} \otimes \sigma_0,
\end{eqnarray}
where $\mathbf{a}(\bk)$ and $\boldsymbol{\lambda}$ are $n^2$-dimensional vectors.

The gap matrix can similarly be written as:
\begin{eqnarray}
\Delta(\bk) = \sum_j \bold{D}_{j}(\bk) \cdot \boldsymbol{\lambda} \otimes \sigma_j (i\sigma_2).
\end{eqnarray}
For a singlet superconductor $\bold{D}_{j}(\bk) $ is an even function in $\bk$ and the index $j=0$. For triplet superconductors $\bold{D}_{j}(\bk) $ is an odd function in $\bk$ and the index $j=\{x,y,z\}$. Each $\bold{D}_{j}(\bk) $ is an $n^2$-dimensional vector with components that carry information about the magnitude, phase and momentum structure of the superconducting order parameter for each orbital. For both the Hamiltonian and gap matrix, the intra-orbital physics is carried by a set of $n$ parameters, related to the $n-1$ Cartan matrices plus the identity, which are all diagonal and with real entries. The remaining parameters correspond to inter-orbital or symmetry breaking effects.

For the determination of $T_c$, starting from the gap equation (Eq.~\ref{LGap}), now we rotate the problem to the band basis and introduce extra indices to label each block with the corresponding band:
\begin{eqnarray}
&&\Delta^a_{s_1s_2}(\bk) = - T \sum_{\bk',n, \{s'\}} \sum_{b,c}V^{ab}_{s_1s_2s_{1'}s_{2'}}(\bk,\bk') \\ \nonumber && \times \, G^{bc}_{s_{2'}s_{3'}}(\bk',i\omega_n) \Delta^{c}_{s_{3'}s_{4'}}(\bk')G^{cbT}_{s_{4'}s_{1'}}(-\bk',-i\omega_n).
\end{eqnarray}
Here we assign only one index to the gap function $\Delta^a_{s_1s_2}(\bk)$ since we consider only intra-band pairing. In the band basis and in the presence of inversion and time-reversal symmetries the GF is block diagonal, so it would also carry only one index, but here we  already foresee the introduction of perturbations which will mix these bands, so we keep both indexes in $G^{ab}_{s_{1}s_{2}}(\bk,i\omega_n)$. We write the interaction matrix in a spectral decomposition, as for the single band case:
\begin{eqnarray}
V^{ab}_{s_1s_2s_{1'}s_{2'}}(\bk,\bk') = v^{ab}  \hat{\Delta}^{a}_{s_1s_2}(\bk)\hat{\Delta}^{b\dagger}_{s_{1'}s_{2'}}(\bk' ).
\end{eqnarray}

With the definitions above the gap equation simplifies to:
\begin{eqnarray}\label{GapMO}
\phi^ a &=& - T\sum_{\bk,n, \{s\}} \sum_{b,c} v^{ab} \phi^{b*}\frac{|\phi^c|}{|\phi^b|} \hat{\Delta}^{b\dagger}_{s_{1}s_{2}}(\bk) \\ \nonumber &&\times \, G^{bc}_{s_{2}s_{3}}(\bk,i\omega_n)  \hat{\Delta}^{c}_{s_{3}s_{4}}(\bk)G^{cbT}_{s_{4}s_{1}}(-\bk,-i\omega_n).
 \end{eqnarray}

The rotation to the band basis introduces great simplification. Once we expand the GF in the perturbation $\delta H(\bk)$, we can write the traces in terms of the non-interacting and non-perturbed GF which has a block diagonal structure with the blocks proportional to the identity, so these can be removed from the trace. A detailed analysis of the gap equation for the determination of the change in $T_c$ for multi-orbital systems is performed in Appendix~\ref{AppTCMO}. Following the lines in the discussion for the single band case, one finds:
\begin{eqnarray}\label{GapMO2}
\phi^{a} &=&  -2 \sum_{b} v^{ab}  N_b(0)  \phi^{b*}  ln \left(\frac{2 e^\gamma}{\pi}\frac{ \omega_c}{T_c}\right) \\ \nonumber &+&   \frac{7 \xi(3)}{32 \pi^2} \sum_{b,c} v^{ab}N_b (0) \phi^{b*}\frac{|\phi^c|}{|\phi^b|}   \frac{ \left\langle Tr[F_{bc}^\dagger F_{bc}] \right\rangle_{FS}}{T_c^2}.
\end{eqnarray}

From the equation above one can quantify the suppression of the critical temperature for multi-band systems. An important feature of this result is that the effects of symmetry breaking fields appear as in the single band case through the \emph{superconducting fitness} $F(\bk)$.

%%%%%%%%%%%%%%%%%%%%%%%%%%%%%%%%%%%%%%%%%
\section{Application to S\MakeLowercase{r}$_2$R\MakeLowercase{u}O$_4$}\label{SRO}

\SRO is a layered prerovskite system and has a strongly 2-dimensional electronic structure (for a review on this material see Mackenzie and Maeno\cite{Mac}). The electrons around the Fermi surface come mostly from the d-orbitals in the $Ru^{4+}$ ions, which are in an octahedral environment. Crystal electric fields split the $4d$ states into two multiplets, with a low lying 3-fold degenerate multiplet formed by $|{yz}\rangle$, $|{xz}\rangle$ and $|{xy\rangle}$ orbitals. For a more economical notation here we label these as $|{1}\rangle$, $|{2}\rangle$ and $|{3}\rangle$ orbitals, respectively.

Based on the symmetries of the orbitals and the underlying lattice, one can construct a tight-binding model with hopping up to next-nearest neighbours and SOC. Introducing the six-dimensional Nambu basis:
\begin{eqnarray}\label{SROBasis}
\Psi\dg_{\bk} = (c\dg_{1\bk\up}, c\dg_{1\bk\down}, c\dg_{2\bk\up},c\dg_{2\bk\down}, c\dg_{3\bk\up}, c\dg_{3\bk\down}),
\end{eqnarray} 
the matrix form of the Hamiltonian is:
\begin{eqnarray}
H_0=\sum_{\bk} \Psi\dg_{\bk}
H_{SRO}(\bk)
\Psi_{\bk},
\end{eqnarray}
with
\begin{eqnarray}
H_{SRO}(\bk)=
\begin{pmatrix}
\epsilon_{1\bk} \sigma_0& t_\bk I_2 +i \eta \sigma_3 & -i \eta \sigma_2\\
t_\bk I_2 -i \eta \sigma_3 & \epsilon_{2\bk} \sigma_0 & i \eta \sigma_1\\
i \eta \sigma_2 & - i \eta \sigma_1& \epsilon_{3\bk} \sigma_0
\end{pmatrix}
\end{eqnarray}
where $\epsilon_{n\bk}$ is the dispersion for each of the orbitals originated from intra-orbital hopping, $t_\bk$ concerns inter-orbital hopping and $\eta$ is the magnitude of the SOC. The explicit form of the dispersions considering up to next nearest neighbour hopping is:
\begin{eqnarray}
\epsilon_{1\bk} &=& -2t_1 \cos k_y-2 t_1'\cos k_x ,\\ \nonumber
\epsilon_{2\bk} &=& -2t_1 \cos k_x -2 t_1'\cos k_y ,\\ \nonumber
\epsilon_{3\bk} &=& -2t_3 (\cos k_x + \cos k_y)-4 t_3'\cos k_x \cos k_y ,\\ \nonumber
t_{\bk} &=& -4t_4 \sin k_x \sin k_y,
\end{eqnarray}
where $t_1$ and $t_1'$ are the hopping amplitudes for intra orbital processes in orbitals $|{1}\rangle$ and $|{2}\rangle$ for neighbours in the direction of and perpendicular to the bond, respectively; $t_3$ and $t_3'$ are the first and second intra-orbital hopping parameters for orbital $|{3}\rangle$; and $t_4$ is the amplitude for inter-orbital hopping to second nearest neighbours between orbitals $|{1}\rangle$ and $|{2}\rangle$. The most important feature of the Hamiltonian above is its structure, which is determined by the character of the low lying orbitals and lattice symmetry. By symmetry, inter-orbital hopping is allowed only between orbitals $|{1}\rangle$ and $|{2}\rangle$, and the presence of SOC introduces mixing of all orbitals.

This matrix form of the Hamiltonian can be concisely written as:
\begin{eqnarray}\label{HSRO}
H_{SRO}(\bk)&=&  \bold{a}(\bk) \cdot \boldsymbol{\lambda}\otimes \sigma_0
\\ \nonumber&+&\eta \left(  \lambda_5\otimes \sigma_2 \!- \!\lambda_2\otimes \sigma_3 \!-\! \lambda_7\otimes \sigma_1\right),
\end{eqnarray}
in terms of direct products of Gell-Mann matrices $\lambda_i$ (the explicit form of the Gell-Mann matrices and some of their properties are summarized in Appendix \ref{GMM}) and Pauli matrices $\sigma_i$. Here $\boldsymbol{\lambda} = (\lambda_0,\lambda_{1},...,\lambda_8)$ is a vector formed by the eight Gell-Mann matrices and $\lambda_0$ is the three dimensional identity matrix. The parameters of the Hamiltonian are summarized in the vector $\bold{a}(\bk)=(a_{0\bk}, a_{1\bk},...,a_{8\bk})$, with
\begin{eqnarray}\label{as}
a_{0\bk}&=&\frac{\epsilon_{1\bk}+\epsilon_{2\bk}+\epsilon_{3\bk}}{3},\\
a_{1\bk}&=&t_\bk,\nonumber\\
a_{3\bk}&=&\frac{\epsilon_{1\bk}-\epsilon_{2\bk}}{2},\nonumber\\
a_{8\bk}&=&\frac{\epsilon_{1\bk}+\epsilon_{2\bk}-2\epsilon_{3\bk}}{2\sqrt{3}},\nonumber
\end{eqnarray} 
and all other $a_{i\bk}=0$ in absence of external fields.

The gap matrix can also be written in the six-dimensional basis introduced in Eq.~\ref{SROBasis}. The most general gap matrix can be written as:
\begin{eqnarray}
\Delta(\bk) = \sum_j \bold{D}_{j} (\bk) \cdot \boldsymbol{\lambda} \otimes \sigma_j (i\sigma_2).
\end{eqnarray}
In analogy to the basis functions $\mathbf{\Psi}^\Gamma(\bk)$ introduced in section~\ref{Val}, here the index $j=0$ refers to a singlet state and $j=\{x,y,z\}$ refers to the three components in the standard d-vector parametrization of the order parameter for triplet SC. Due to the presence of different orbitals here $\bold{D}_{j} (\bk) $ also carries information about the different order parameter magnitudes. $\bold{D}_{j} (\bk)=(D_{j0\bk}, D_{j1\bk},...,D_{j8\bk})$ is a 9-dimensional vector with components:
\begin{eqnarray}
D_{j0\bk}&=&\frac{ \phi_1d_{j\bk}^1+\phi_2d_{j\bk}^2+\phi_3d_{j\bk}^3}{3},\\
D_{j2\bk}&=&i \phi_{12}d_{j\bk}^{12},\nonumber\\
D_{j3\bk}&=&\frac{\phi_1d_{j\bk}^1-\phi_2d_{j\bk}^2}{2},\nonumber\\
D_{j5\bk}&=&i \phi_{13}d_{j\bk}^{13},\nonumber\\
D_{j7\bk}&=&i \phi_{23}d_{j\bk}^{23},\nonumber\\
D_{j8\bk}&=&\frac{\phi_1d_{j\bk}^1+\phi_2d_{j\bk}^2-2\phi_3d_{j\bk}^3}{2\sqrt{3}}, \nonumber
\end{eqnarray} 
with all other $D_{ji\bk}=0$. Here $d_{j\bk}^a$ represents the j-component of the intra-orbital order parameter in orbital $a$ and $d_{j\bk}^{ab}$ is the j-component of the inter-orbital order parameter between orbitals $a$ and $b$.

%%%%%%
\subsection{Stability of the SC gap}\label{Sta}

Here we want to study the stability of the SC gap structure for \SRO based on the \emph{superconducting fitness} $F(\bk)$ which gives a measure of the incompatibility of the SC state and the underlying electronic structure. Inter-orbital pairing always gives rise to non-zero contributions to $F(\bk)$, indicating that these are not very robust SC states. We start by analyzing $F(\bk)$ in the absence of external symmetry breaking fields and inter-orbital pairing. We show below that the  \emph{superconducting fitness} gives different results for distinct pairing states, due to the non-trivial underlying orbital structure and lattice symmetry of \SRO\!, and from those we can argue towards the most stable pairing state for this material.

In order to understand the effects of inter-orbital hopping and SOC in the determination of the gap structure, we consider four different SC states, omitting the momentum dependence:
\begin{eqnarray}
\Delta_{S}&=&  \bold{D}_0 \cdot \boldsymbol{\lambda} \otimes  (i\sigma_2), \\ \nonumber
\Delta_{Tx}&=& - \bold{D}_x \cdot \boldsymbol{\lambda} \otimes \sigma_3,\\ \nonumber
\Delta_{Ty} &=&  i \bold{D}_y\cdot \boldsymbol{\lambda} \otimes \sigma_0,\\ \nonumber
\Delta_{Tz} &=& \bold{D}_z\cdot \boldsymbol{\lambda} \otimes \sigma_1,
\end{eqnarray}
a singlet state and triplet states with d-vector along the x,y and z-axis, respectively. 

In absence of inter-orbital hopping and SOC ($t_4=0$ and $\eta=0$), we have three decoupled orbitals and $F=0$. In this case SC develops independently in each orbital and the critical temperature $T_c \sim e^{-1/2N_a(0)v_a}$ is determined by the largest value of $N_a(0)v_a$, where $a=\{1,2,3\}$ is the orbital index. 

In \SRO inter-orbital effects are small and can be treated as a perturbation. One can see the effect of inter-orbital hopping (IOH) by turning on $t$ and evaluating the \emph{superconducting fitness} to find:
%\begin{eqnarray}
%F_{IOH}&=& i t (\hat{\Phi}_{1}-\hat{\Phi}_{2}) M_{20}\\ \nonumber
%&=& i t(\hat{d}_{x1}-\hat{d}_{x2}) M_{21}\\ \nonumber
%&=& i t(\hat{d}_{y1}-\hat{d}_{y2}) M_{22}\\ \nonumber
%&=& i t(\hat{d}_{z1}-\hat{d}_{z2}) M_{23}
%\end{eqnarray}
\begin{eqnarray}
F_{IOH}&=& i t _4(\hat{\phi}^1d_{j}^1-\hat{\phi}^2d_{j}^2) M_{2j}
\end{eqnarray}
for any SC state. Here we introduced the short notation $M_{ij}=\lambda_i \otimes \sigma_j$. Note that for all SC states $F$ is zero only if the gaps in orbitals $|{1}\rangle$ and $|{2}\rangle$ have the same phase and momentum structure.

In similar lines one can analyze the effect of SOC in absence of inter-orbital hopping, evaluating the \emph{superconducting fitness} for the different SC states above to find:
\begin{eqnarray}\label{SOCs}
F_{SOC,S}&=& i \eta (\hat{\phi}^1 d_0^1-\hat{\phi}^2d_0^2)M_{13} \\ \nonumber&+& i \eta (\hat{\phi}^2d_0^2-\hat{\phi}^3d_0^3)M_{61} - i \eta (\hat{\phi}^1d_0^1-\hat{\phi}^3d_0^3)M_{42}
\end{eqnarray}
for $\Delta_{S}$, indicating that SOC tends to lock the gaps in all orbitals to have the same phase momentum structure;
\begin{eqnarray}\label{SOCx}
F_{SOC,T_x}&=&i  \eta (\hat{\phi}^1d_{x}^1+\hat{\phi}^2d_{x}^2)M_{22} \\ \nonumber&+& i \eta (\hat{\phi}^2d_{x}^2-\hat{\phi}^3d_{x}^3)M_{60} + i \eta (\hat{\phi}^1d_{x}^1+\hat{\phi}^3d_{x}^3)M_{53}
\end{eqnarray}
for $\Delta_{Tx}$, in which case $T_c$ is optimized for $-\hat{\phi}^1d_{x}^1=\hat{\phi}^2d_{x}^2=\hat{\phi}^3d_{x}^3$;
\begin{eqnarray}\label{SOCy}
F_{SOC,T_y}&=& - i  \eta (\hat{\phi}^1d_{y}^1+\hat{\phi}^2d_{y}^2)M_{21} \\ \nonumber&+& i \eta (\hat{\phi}^2d_{y}^2+\hat{\phi}^3 d_{y}^3)M_{73} - i \eta (\hat{\phi}^1d_{y}^1-\hat{\phi}^3d_{y}^3)M_{40}
\end{eqnarray}
for $\Delta_{Ty}$, with optimized state for $\hat{\phi}^1d_{y}^1=-\hat{\phi}^2d_{y}^2=\hat{\phi}^3d_{y}^3$; and
\begin{eqnarray}\label{SOCz}
F_{SOC,T_z}&=&i  \eta (\hat{\phi}^1d_{z}^1-\hat{\phi}^2d_{z}^2)M_{10} \\ \nonumber&-& i \eta (\hat{\phi}^2d_{z}^2+\hat{\phi}^3d_{z}^3)M_{72} - i \eta (\hat{\phi}^1d_{z}^1+\hat{\phi}^3d_{z}^3)M_{51}
\end{eqnarray}
for $\Delta_{Tz}$, in which case $\hat{\phi}^1d_{z}^1=\hat{\phi}^2d_{z}^2=-\hat{\phi}^3d_{z}^3$ leads to the most favourable SC state.

Note that if both IOH and SOC are present, there is no triplet state configuration with an in-plane d-vector which is in accordance with the underlying electronic structure (or that satisfy $F_{IOH}=0$ and $F_{SOC}=0$ concomitantly). For a singlet and a triplet state with d-vector along the z-axis the gap structure can be made compatible with the underlying electronic properties.  Focusing on the triplet state, this indicates that for the d-vector in the z-direction the structure of the gap is expected to be the same for all orbitals and that the sign of the gap in the band dominated by the $|3 \rangle$ orbital (usually referred to as the $\gamma$-band) should be the opposite to the sign of the gap in the bands dominated by the $| 1 \rangle$ and $| 2 \rangle$ orbitals ($\alpha$ and $\beta$ bands).

In conclusion, based on the analysis of the \emph{superconducting fitness} developed above, we can claim that the most stable SC state is the triplet state with the D-vector along the z-direction (or singlet, which here we neglect given the experimental support for a triplet SC state in \SRO \cite{Mac}) since this kind of order parameter can find a configuration such that it is completely compatible with the electronic structure. While this result is consistent with other theoretical discussions on the effect of SOC we would like to note that with our analysis only the compatibility is tested, but not any other effect of SOC and IOH, for instance, on the pairing interaction. Here we consider only the structure of the bare Hamiltonian and SC state in the multi-orbital Nambu basis without any specification on the origin of the pairing or the fine details of the band structure. What is important for this conclusion is the orbital content of the low lying states and the symmetry of the lattice.

The \emph{superconducting fitness} can be used as a guide for the analysis of the stability of SC states for a given non-interacting Hamiltonian, but it should not be used as a definitive answer for it as it overlooks the pairing mechanism and the effects of enhanced DOS due to the proximity of Van Hove singularities, which might be important for \SRO\!. Next we apply this concept in the analysis of the behaviour of the critical temperature $T_c$ in presence of small symmetry breaking fields. For this kind of analysis, starting from a given SC state, the analysis of the \emph{superconducting fitness} leads to much more powerful arguments.

%%%%%%
\subsection{Time-reversal symmetry breaking effects}\label{TRS}

The presence of external magnetic fields leads to time-reversal symmetry breaking and possible suppression of the SC state as already discussed in Sec.~\ref{Val}. In general, the effects of magnetic fields can appear in two different ways: by direct coupling to the spin degree of freedom as a Zeeman term or by coupling to the orbital angular momentum. For the specific case of \SRO\!\!, we can write the explicit matrix form for these effects with the multi-orbital basis introduced in Eq.~\ref{SROBasis}. We start with the Zeeman term:
%\begin{eqnarray}
%\delta H_{Z} &=& -\mathbf{h}\cdot \lambda_0\otimes \boldsymbol{\sigma}_T\\ \nonumber
%&=& \begin{pmatrix}
% -\mathbf{h}\cdot \boldsymbol{\sigma}_T & 0 & 0\\
% 0 & -\mathbf{h}\cdot \boldsymbol{\sigma}_T & 0\\
%0 & 0 & -\mathbf{h}\cdot \boldsymbol{\sigma}_T
%\end{pmatrix},
%\end{eqnarray}
\begin{eqnarray}
\delta H_{Z} &=& -\mathbf{h}\cdot \lambda_0\otimes \boldsymbol{\sigma}_T
\end{eqnarray}
where $\mathbf{h}=(h_x,h_y,h_z)$ is the external magnetic field. Note that this perturbation is independent of the orbital character.  Focusing only on the field dependent part of the \emph{superconducting fitness}, 
neglecting any intrinsic suppression of SC due to the bare electronic structure, we find
\begin{eqnarray}\label{FZ}
F_{Z,S}&=&2  \begin{pmatrix}
\hat{\phi}^1d_0^1 \mathbf{h}\cdot \boldsymbol{\sigma}_T & 0 & 0\\
0 &\hat{\phi}^2 d_0^2 \mathbf{h}\cdot \boldsymbol{\sigma}_T & 0 \\
0 &  0 &\hat{\phi}^3 d_0^3 \mathbf{h}\cdot \boldsymbol{\sigma}_T
\end{pmatrix} , \\
F_{Z,T}&=&2  \begin{pmatrix}
\hat{\phi}^1 \mathbf{d}^1 \cdot \mathbf{h} \sigma_0& 0 & 0\\
0 & \hat{\phi}^2\mathbf{d}^2 \cdot \mathbf{h} \sigma_0 & 0 \\
0 &  0 & \hat{\phi}^3\mathbf{d}^3 \cdot \mathbf{h} \sigma_0
\end{pmatrix},\nonumber
\end{eqnarray}
for singlet and triplet states, respectively.  So the coupling of magnetic field to the spin degree of freedom leads to the same suppression of the critical temperature as in the single band case. Singlet SC is always suppressed in the presence of magnetic field, while triplet SC is suppressed only if the magnetic field has a component in the same direction as the d-vector.

On the other hand, considering the coupling of the magnetic field to the orbital degree of freedom:
\begin{eqnarray}\label{Horb}
\delta H_{Orb}
&=& -\mathbf{h}_{o}\cdot \boldsymbol{\lambda}\otimes \sigma_0,\\ \nonumber
&=& \begin{pmatrix}
0 & -h_{oz} \sigma_0 & i h_{oy} \sigma_0\\
-h_{oz} \sigma_0 & 0 & -h_{ox} \sigma_0 \\
-i h_{oy} \sigma_0 &  -h_{ox} \sigma_0 &0
\end{pmatrix},
\end{eqnarray}
where $\mathbf{h}_{o}=(h_{oz},0,0,0,h_{oy},h_{ox},0,0)$ and $\boldsymbol{\lambda} = (\lambda_0,\lambda_{1},...,\lambda_8)$. Here the orbital content is important to determine the structure of the matrix $\delta H_{Orb}$ and ultimately the effects of magnetic field on the superconducting state. Computing the \emph{superconducting fitness} in order to verify the stability of different gap structures in the presence of orbital effects, we find the following:
\begin{eqnarray}\label{FOS}
F_{Orb}&=& -i h_{ox} (\hat{\phi}^2 d_{j}^2-\hat{\phi}^3d_{j}^3) M_{7j}\\ \nonumber&+& h_{oy} (\hat{\phi}^1d_{j}^1+\hat{\phi}^3d_{j}^3) M_{5j} + i h_{oz} (\hat{\phi}^1d_{j}^1-\hat{\phi}^2d_{j}^2) M_{2j},
\end{eqnarray}
Here $j=0$ for singlet SC and $j=\{x,y,z\}$ for triplet SC. Note that for any gap structure among the different orbitals it is impossible to have all terms in $F_{Orb}$ equal to zero. Now magnetic field is potentially detrimental to all SC states irrespective of its direction. If we consider the optimal order parameter according to the analysis in the previous subsection, a triplet state with the d-vector along the z-direction such that  $\hat{\phi}^1d_{z}^1=\hat{\phi}^2d_{z}^2=-\hat{\phi}^3d_{z}^3=\hat{\phi} d_z$, we find:
\begin{eqnarray}
F_{Orb}&=& -2 i h_{ox} \hat{\phi} d_z M_{73},
\end{eqnarray}
what indicates that now a transverse field is also detrimental to a SC state with d-vector along the z-direction. We can also consider a sub-optimal order parameter, a triplet state with the d-vector along the x-direction, which in case of strong SOC will tend to satisfy $-\hat{\phi}^1d_{x}^1=\hat{\phi}^2d_{x}^2=\hat{\phi}^3d_{x}^3=\hat{\phi} d_x$. In this case:
\begin{eqnarray}
F_{Orb}&=& 2 i h_{oz} \hat{\phi} d_x M_{71} ,
\end{eqnarray}
indicating that now magnetic field along the z-axis is also detrimental to a SC state with d-vector in the plane.

From the discussion above one conclude that for multi-orbital systems the susceptibility of the critical temperature in presence of magnetic fields is not trivial and goes beyond the standard discussion of limiting effects. We call this new effect \emph{inter-orbital effect} (in order to discriminate it from the usual orbital effect) and explore its implications on our understanding of recent experiments on \SRO in section \ref{Discussion}.

%%%%%%
\subsection{Inversion symmetry breaking effects}

In an analogous fashion we can analyse the \emph{superconducting fitness} to determine the stability of SC states under inversion symmetry breaking. The lack of inversion symmetry can be implemented in the Hamiltonian by a Rashba-type SOC term, as discussed in the single band scenario above:
%\begin{eqnarray}
%\delta H_{Inv} &=& \mathbf{g} (\bk) \cdot \lambda_0\otimes \boldsymbol{\sigma}_T,\\ \nonumber
%&=& \begin{pmatrix}
%\mathbf{g}(\bk)\cdot \boldsymbol{\sigma}_T & 0 & 0\\
 %0 & \mathbf{g}(\bk)\cdot \boldsymbol{\sigma}_T & 0\\
%0 & 0 & \mathbf{g}(\bk)\cdot \boldsymbol{\sigma}_T
%\end{pmatrix},
%\end{eqnarray}
\begin{eqnarray}
\delta H_{Inv} &=& \mathbf{g} (\bk) \cdot \lambda_0\otimes \boldsymbol{\sigma}_T,
\end{eqnarray}
where $\mathbf{g}(\bk)=(g_x(\bk),g_y(\bk),g_z(\bk))\sim  \xi \left(\hat{i}\times \bk\right)$ is an odd function in $\bk$, $\hat{i}$ defines the direction of the external symmetry breaking field and $\xi$ is a coupling constant. This is the analogous of the Zeeman perturbation for time-reversal symmetry breaking, as it brings only intra-orbital effects. For simplicity, here we assume inversion symmetry breaking has the same effect in all orbitals. In this case the \emph{superconducting fitness} gives:
\begin{widetext}
\begin{eqnarray}\label{FInv}
F_{Inv,S}&=&0, \\
F_{inv,T}&=&-2 i   \begin{pmatrix}
\hat{\phi}^1(\mathbf{g}\times \mathbf{d}^1) \cdot \boldsymbol{\sigma}_T& 0 & 0\\
0 &\hat{\phi}^2(\mathbf{g}\times  \mathbf{d}^2) \cdot \boldsymbol{\sigma}_T& 0 \\
0 &  0 &\hat{\phi}^3(\mathbf{g}\times  \mathbf{d}^3) \cdot \boldsymbol{\sigma}_T
\end{pmatrix}.\nonumber
\end{eqnarray}
\end{widetext}
Due to the intra-orbital character of this perturbation the single-band result for the reduction of the critical temperature can be directly generalized for the multi-orbital case: singlet SC is not susceptible in the presence of inversion symmetry breaking, while triplet SC is suppressed only if $\mathbf{g}(\bk)$ has a component perpendicular the d-vector.

Orbital effects are also important when analysing the effects of inversion symmetry breaking. The analogous perturbation to the orbital effect discussed for time reversal symmetry breaking is:
\begin{eqnarray}
\delta H_{InvOrb} &=& \mathbf{g}_{o}(\bk)\cdot \boldsymbol{\lambda}\otimes \sigma_0,\\ \nonumber
&=& \begin{pmatrix}
0 & g_{oz}(\bk) \sigma_0 & -i g_{oy}(\bk) \sigma_0\\
g_{oz}(\bk) \sigma_0 & 0 & g_{ox} (\bk) \sigma_0 \\
i g_{oy} (\bk) \sigma_0 &  g_{ox} (\bk) \sigma_0 &0
\end{pmatrix},
\end{eqnarray}
where $\mathbf{g}_{o}(\bk)=(g_{oz}(\bk),0,0,0,g_{oy}(\bk),g_{ox}(\bk),0,0)$. Analyzing the\emph{superconducting fitness} in order to verify the stability of different gap structures in the presence of orbital effects, we find, omitting the momentum dependence:
\begin{eqnarray}\label{FInvO}
F_{InvOrb}&=& - g_{ox} (\hat{\phi}^2d_{j}^2+\hat{\phi}^3d_{j}^3) M_{6j}\\ \nonumber&-& i g_{oy} (\hat{\phi}^1d_{j}^1-\hat{\phi}^3d_{j}^3) M_{4j} + g_{oz} (\hat{\phi}^1d_{j}^1+\hat{\phi}^2d_{j}^2) M_{1j}.
\end{eqnarray}
Note that $F_{InvOrb}$ cannot be zero for any gap structure, so we expect inversion symmetry breaking to be potentially detrimental for both singlet and triplet SC states, irrespective of the relative direction of the symmetry breaking field and the d-vector for the triplet case.

Now we analyse some cases of interest for \SRO\!\!. For the optimal order parameter, d-vector along the z-direction with $\hat{\phi}^1d_{z}^1=\hat{\phi}^2d_{z}^2=-\hat{\phi}^3d_{z}^3=\hat{\phi} d_z$:
\begin{eqnarray}
F_{InvOrb}&=& - 2i \hat{\phi} d_z g_{oy} M_{43} + 2 \hat{\phi} d_z g_{oz} M_{13},
\end{eqnarray}
so now inversion symmetry breaking is also detrimental for the field $\mathbf{g}(\bk)$ in the direction of the d-vector. For the sub-optimal order parameter, a triplet state with the d-vector in the x-direction $-\hat{\phi}^1d_{x}^1=\hat{\phi}^2d_{x}^2=\hat{\phi}^3d_{x}^3=\hat{\phi} d_x$:
\begin{eqnarray}
F_{InvOrb}&=& 2\hat{\phi} d_x  g_{ox}  M_{61} - 2i\hat{\phi} d_x g_{oy} M_{41},
\end{eqnarray}
what also indicates that now inversion symmetry breaking is detrimental irrespective of its direction, a direct consequence of \emph{inter-orbital effects}. 

%%%%%%
\subsection{Discussion}\label{Discussion}

%d-vector in the z-direction
The importance of SOC in the determination of the direction of the d-vector in \SRO is a well stablished result discussed by several authors\cite{SigNg, Yan, Sca14}.  While the analysis of the \emph{superconducting fitness} cannot address the specific momentum dependence of the gap (beyond favouring the state with maximal condensation energy), it is in agreement with previous results finding that the most favourable triplet state has the d-vector along the z-direction. We would like to emphasize here that our analysis, based solely on the modified commutation of the Hamiltonian and gap matrixes, gives an important new insight to this issue from another viewpoint . We would also like to highlight that SOC is not alone responsible, but that its interplay with inter-orbital hopping is essential as well (see discussion in Sec.~\ref{Sta}).
%Ultimately, the criterium considers the structure of the SOC and inter-orbital hopping in the multi-orbital Nambu basis, which is dictated by the orbital angular momentum of the electrons and the lattice symmetry.

%Importance of three bands
The need for a complete treatment considering all three orbitals in presence of SOC also becomes evident from the analysis of the \emph{superconducting fitness}.  Superconductivity being present in all three bands is essential for the fitness as can be easily verified examining $ F(\bk)$ in Eqs.~\ref{SOCs}-\ref{SOCz}. %Note that in case of pairing only in the $\gamma$-band (with strong $|xy\rangle$ character, labelled here by $|3\rangle$), $F(\bk)$ gives terms proportional to $d^3 \eta$ (see Eqs.~\ref{SOCs}-\ref{SOCz}) and it is impossible to make superconductivity completely compatible with the electronic structure. On the other hand, if we also have pairing in the other two orbitals, we can actually make SC stable from the perspective of the \emph{superconducting fitness}.
% Considering pairing only in the $\alpha$ and $\beta$-bands (with strong $|yz\rangle$ and $|xz\rangle$ character, labelled here by $|1\rangle$ and $|2\rangle$, respectively), in an attempt to work with a simplified two-orbital model for \SRO\!\!, one is able to find a compatible gap configuration, again with the d-vector in the z-direction. The problem with this simplification is the fact that it does not allow a proper description of \emph{inter-orbital effects}, as will be discussed in detail below. For this reason we put forward the effort of discussing the full three-orbital model.

%Magnetic field effects
Concerning magnetic field effects in \SRO\!\!, experiments show two features that are still not completely understood. The first concerns the anisotropy of the upper critical field, which does not follow a pure effective mass model as expected for an ordinary layered type-II superconductor subject to standard orbital depairing\cite{Deg,Yon13,Yon14}. Secondly, for fields within $2^o$ from the plane the transition is actually first order at temperatures below $0.8K$, as recently observed by magnetocaloric effect\cite{Yon13} and specific heat measurements in ultra pure samples\cite{Yon14}. The presence of a first order transition suggests that the Pauli limiting effect is at work \cite{Ama15}, but this would be inconsistent with NMR\cite{Ish98,Ish01,Mur} and neutron scattering\cite{Duf} experiments which show no observable change in the spin susceptibility through the superconducting transition. An explanation for this unusual behaviour, compatible with all experimental results,  has not been established so far, and experimentalists have been suggesting that it is due to a new pair breaking mechanism\cite{Yon14,Deg}. Here we suggest that the mechanism could be related to \emph{inter-orbital effects} (IOE), which can potentially account for \emph{both} the anisotropy and the change in the character of the transition, as we discuss in the following.

In type-II SC with large coherence length, orbital effects are usually the dominant pair breaking mechanism and the suppression of the SC state occurs in a second order phase transition\cite{Tin}.  In layered materials, anisotropies in the effective mass are important and lead to anisotropies in the upper critical field, usually described by an effective mass model (EMM)\cite{Mor}. Following the notation introduced in Yonezawa et al.\cite{Yon13} the angular dependence of the upper critical field can be written as:
\begin{eqnarray}\label{EMM}
h_{c2}^{EMM}(\theta) = \frac{h_{c2}(90^o)}{\sqrt{\sin^2\theta + \cos^2\theta/\Gamma^2 }},
\end{eqnarray}
where $\Gamma$ is the anisotropy parameter. As can be seen in Fig.~\ref{Angle}, the EMM (blue, full line) describes the angular dependence of the upper critical field very well for angles larger than $2^o$. 

\begin{figure}[t]
\begin{center}
\includegraphics[width=\linewidth, keepaspectratio]{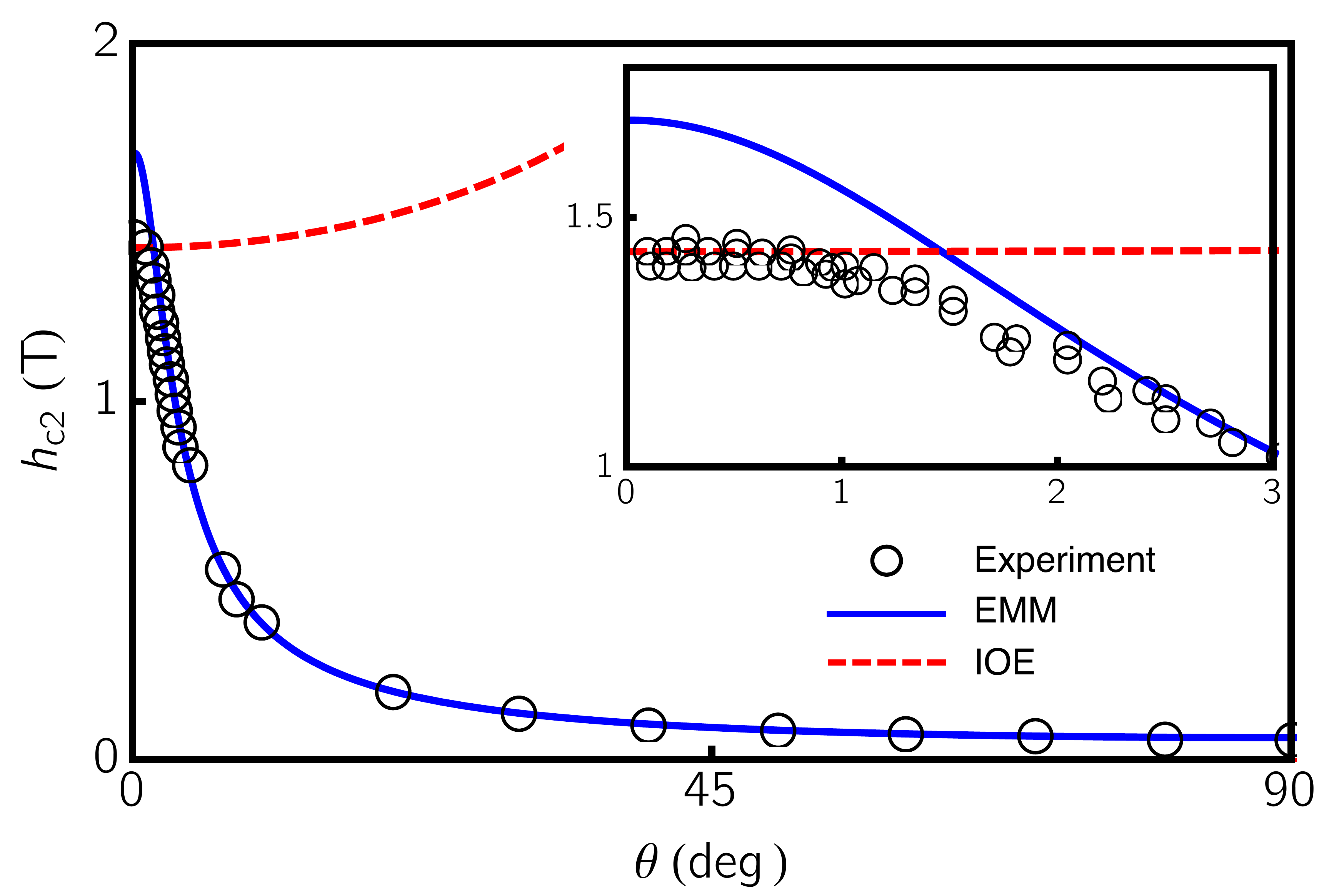}\label{Angle}
\caption{Angular dependence of the upper critical field ($\theta=0^o$ corresponds to fields in plane). The blue (full) line is the best fit for angles $\theta>2^o$ with the effective mass model (EMM) of Eq.~\ref{EMM} with parameters $\Gamma=25$ and $h_{c2}(90^o)=0.07 T$.  The red (dashed) line refers to the inter-orbital effect (IOE) plotted phenomenologically as $h_{c2}^{IOE}= \frac{h_{c2}(0^o)}{\cos\theta}$, with $h_{c2}(0^o)=1.44 T$. The circles are experimental data from Yonezawa et al.\cite{Yon13}. The inset shows the small angle region.}
\label{Angle}
\end{center}
\end{figure}

The picture changes for angles within $2^o$ of the plane. For this range of angles the standard orbital mechanism cannot explain the observations since there are clear deviations from the EMM and concomitantly the transition becomes first order\cite{Deg,Yon13,Yon14}. The fact that $h_{c2}$ is smaller in this region indicates that there should be an extra mechanism which suppresses SC which is most effective for in-plane fields. Standard paramagnetic limiting effects cannot account for the observations since there should be no suppression of SC for in-plane fields once we assume the d-vector is along the z-direction.

Due to the multi-orbital character of \SRO\!\!, one should also consider the coupling of the magnetic field to the orbital angular momentum. From the analysis of the \emph{superconducting fitness} in presence of such coupling (see Sec.~\ref{TRS}), we learned that an in-plane magnetic field can suppress SC even for a d-vector along the z-direction (see Eq.~\ref{FOS} and subsequent discussion), in clear contrast to the standard paramagnetic limiting effect. Now the orbital polarization of states within the same band is harmful to superconductivity, in analogy to the standard paramagnetic effect in which case the spin polarization is detrimental to the SC state.

From a  free energy analysis (similar to the standard analysis for the paramagnetic limiting effect), we can find an estimate of the change in the orbital susceptibility necessary for the inter-orbital mechanism to take place. This is of course a simplified analysis since we are effectively looking at a single-order-parameter (''single-band'') description. Given a Landau expansion for the free energy in presence of a magnetic field in the x-direction coupling to the orbital DOF:
\begin{eqnarray}
F=a(T)|\Delta|^2+ b|\Delta|^4 - \frac{\chi_{Orb} h_x^2}{2},
\end{eqnarray}
where $|\Delta|$ is the magnitude of the order parameter and 
\begin{eqnarray}
a(T)=\frac{N(0)}{T_c^0}(T-T_c^0), \hspace{0.5cm} b=\frac{7\xi(3)}{16 \pi^2}\frac{N(0)}{(T_c^0)^2},
\end{eqnarray}
are the standard Landau free energy parameters, with $N(0)$ the DOS and $T_c^0$ the critical temperature of the unperturbed system. Here $\chi_{Orb}$ is the orbital susceptibility in the normal phase. We can now use the result on the reduction of the critical temperatures for a correction to the first term in the free energy in presence of external symmetry breaking fields. Considering the d-vector along the z-direction and the magnetic field in the x-direction we find:
\begin{eqnarray}
a(T)\rightarrow \frac{N(0)}{T_c^0}\left(T-T_{c}^0 + \frac{7\xi(3)}{16 \pi^2}\frac{h_x^2}{( T_c^0)^2}\right).
\end{eqnarray}
From this correction to the free energy we can identify the change in the orbital susceptibility from the normal state to the superconducting state at zero temperature:
\begin{eqnarray}
\Delta \chi_{Orb} = 2 N(0) \frac{7\xi(3)}{16 \pi^2} \frac{|\Delta|^2}{T_c^2}.
\end{eqnarray}
Using the result of the minimization of the free energy $|\Delta|^2 = - a(T)/2b$, we find $\Delta \chi_{Orb}=N(0)$ in rough extrapolation to zero temperature, similar to the standard result for the paramagnetic limiting effect.

If the transition due to IOE is of first order, the upper critical field is determined by comparing the magnetic and condensation energies. The condensation energy can be estimated as:
\begin{eqnarray}
E_C=-\frac{N(0) |\Delta|^2}{2}.
\end{eqnarray}

The magnetic energy now is related to the polarizarion of the orbitals within a given band, so here we use $\Delta\chi_{Orb}$ as the change in the orbital susceptibility from the normal to the SC state:
\begin{eqnarray}
E_M=-\frac{\Delta\chi_{Orb}h_x^2}{2},
\end{eqnarray}
which lead to the critical field
\begin{eqnarray}
h_{c2}^{IOE} \approx \sqrt{\frac{N(0)}{\Delta\chi_{Orb}}} \frac{T_{c}^0}{\cos\theta},
\end{eqnarray}
where $\theta$ is the angle the magnetic field makes with the conducting plane.

For a test we can compare the critical fields from the EMM and from the IOE. From experiments we know that the crossing of the EMM and the IEO effects should happen at approximately $\theta=2^o$, so at this angle we would expect:
\begin{eqnarray}
 \sqrt{\frac{N(0)}{\Delta\chi_{Orb}}} \frac{T_{c}^0}{\cos 2^o} = h_{c2}^{EMM}(2^o),
\end{eqnarray}
plugging in the experimental values ($T_{c}^0 \sim 1.5K$ and $h_{c2}^{EMM}(2^o)\sim 1.3 T$) we find $\Delta \chi_{Orb}\sim N(0)$, what means that the change in the orbital susceptibility from the normal to the superconducting state should be of the order of the DOS, consistent with the free energy analysis above. 

Beyond the crude estimates, here we would like to highlight the mechanism, the \emph{inter-orbital effect}, as a possible path to suppress superconductivity in multi-orbital systems. For \SRO\!\!, note that only transverse (or in-plane) magnetic fields can polarize the orbital part. From Eqs.~\ref{HSRO}, \ref{as} and \ref{Horb} one can see that a field in the z-direction only renormalizes the inter-orbital hopping parameter $t_\bk$, not leading to orbital polarization. In contrast to the standard orbital depairing leading to the EMM, which is more effective for fields along the z-direction, the \emph{inter-orbital effect} can be stronger for fields in the plane. The presence of these two mechanisms will lead to an interplay, with the standard orbital suppression dominating for large angles leading to a second order phase transition and the  \emph{inter-orbital effect} dominating for small angles and potentially leading to a first order phase transition (in analogy to the standard paramagnetic limiting effect), as can be seen in Fig.~\ref{Angle}. Therefore, the \emph{inter-orbital effect} may account for both puzzling features in the upper critical field of \SRO. 

Analogously, multi-orbital effects qualitative change our understanding of the stability of superconducting states in presence of inversion symmetry breaking fields, something that should be further explored both theoretically and experimentally.

%%%%%%%%%%%%%%%%%%%%%%%%%%%%%%%%%%%%
\section{Conclusion}\label{Con}

%Criterium
In this paper we propose a general framework to test different pairing states against a given non-interacting Hamiltonian in order to find a SC phase compatible with the basic electronic structure of the normal state. We introduce the concept called \emph{superconducting fitness} which is based on a modified commutation relation of the bare Hamiltonian and the gap matrix in the appropriate basis. We validate this idea with well-known results from weak coupling theories in presence of fields breaking key symmetries, and find that it gives a direct measure of the suppression of the critical temperature in case it is non-zero.

%SRO: d-vector direction and inter-orbital effects
We analyse the \emph{superconducting fitness} for the multi-orbital system \SRO\!\!. The analysis indicates that in presence of both inter-orbital hopping and SOC the most compatible SC state favors the d-vector along the z-axis.  Also, from the criterium, we are able to have a first understanding of \emph{inter-orbital effects} in the suppression of different SC states, based on which we propose an explanation for the unusual behaviour of the upper critical field of \SRO\!\!.

%Perspectives
We believe this work brings in some new insights for the general problem of conventional and unconventional superconductivity in multi-orbital systems and that is can be applied to other materials and model systems. It motivates a more quantitative microscopic analysis of the \emph{inter-orbital effects} in \SRO in order to confirm the ideas above concerning magnetic field effects. The consequences for inversion symmetry breaking should also be reviewed in the light of the new \emph{inter-orbital effects}. The concept of \emph{fitness} also suggests directions for future work which include the proposal of similar analysis for other kinds of order.

%%%%%%%%%%%%%%%%%%%%%%%%%%%%%%%%%%%%%%%%%
\begin{acknowledgements}
We thank M.H. Fischer for stimulating discussions. This work was supported by Dr. Max R\"{o}ssler, the Walter Haefner Foundation and the ETH Zurich Foundation (AR) and by the Swiss National Science Foundation (MS).
\end{acknowledgements}

%%%%%%%%%%%%%%%%%%%%%%%%%%%%%%%%%%%%%%%%%
\appendix
\section{Derivation of the relation of $F(\bk)$ with the suppression of $T_c$}\label{AppTC}

In this appendix we give the details of the derivation of Eq.~\ref{TC}. We start with the linearized gap equation:
\begin{eqnarray}
1=&& - Tv  \sum_{\bk,n}Tr \left[  \hat{\Delta}^\dagger(\bk) G(k)  \hat{\Delta}(\bk)G^T(-k)\right],
\end{eqnarray}
where we use the notation $G(k)=G(\bk,i\omega_n)$, as in the main text. Writing
\begin{eqnarray}\label{ExpdH}
G(k) &=& G_0(k)  \sum_{p=0}^\infty  \left( G_0(k)\delta H(\bk)\right)^p,
\end{eqnarray}
keeping only terms up to second order in $\delta H(\bk)$, introducing the even shorter notation $ G_0=G_0(\bk,i\omega_n)$ and $ \bar{G}_0=G_0(-\bk,-i\omega_n)$, we find:
\begin{eqnarray}\label{EqParts}
1&=& - Tv  \sum_{\bk,n} \left\{ G_0\bar{G}_0 T_1+ \left(G_0\right)^2\bar{G}_0T_2 +  G_0\left(\bar{G}_0\right)^2 T_3 \right. \nonumber\\&+&\left.  \left(G_0\bar{G}_0\right)^2 T_4+ \left(G_0\right)^3\bar{G}_0T_5+  G_0\left(\bar{G}_0\right)^3 T_6\right\},\nonumber
\end{eqnarray}
where
\begin{eqnarray}
T_1&=&Tr \left[  \hat{\Delta}^\dagger(\bk)  \hat{\Delta}(\bk)\right],\\\nonumber
T_2&=&Tr \left[  \hat{\Delta}^\dagger(\bk) \delta H (\bk)  \hat{\Delta}(\bk)\right],\\\nonumber
T_3&=&Tr \left[  \hat{\Delta}^\dagger(\bk) \hat{\Delta}(\bk)[\delta H(-\bk)]^T\right],\\\nonumber
T_4&=&Tr \left[  \hat{\Delta}^\dagger(\bk) \delta H (\bk)  \hat{\Delta}(\bk)[\delta H(-\bk)]^T\right],\\\nonumber
T_5&=&Tr \left[  \hat{\Delta}^\dagger(\bk) \left(\delta H (\bk)\right)^2  \hat{\Delta}(\bk)\right],\\\nonumber
T_6&=&Tr \left[  \hat{\Delta}^\dagger(\bk) \hat{\Delta}(\bk)[\left(\delta H(-\bk)\right)^2]^T\right].
\end{eqnarray}
We can rewrite the product:
\begin{eqnarray}
 \hat{\Delta}^\dagger(\bk) \hat{\Delta}(\bk)  =|\Psi^\Gamma(\bk)|^2 I_2 + \mathbf{q}(\bk)\cdot\boldsymbol{\sigma}^T,
\end{eqnarray}
where $ \mathbf{q}(\bk) = i  \left( \Psi^\Gamma(\bk)\right)^*\times  \Psi^\Gamma(\bk)$.

We now analyse each term separately:

\begin{itemize}

\item The first term has the simplest trace:
\begin{eqnarray}
T_1  &=& 2|\Psi^\Gamma(\bk)|^2,
\end{eqnarray}
and the sums over $\bk$ and Matsubara frequencies lead to the standard BCS result, so the first term contributes to the right hand side of the gap equation with:
\begin{eqnarray}
- 2 v N(0) ln \left(\frac{2 e^\gamma}{\pi}\frac{ \omega_c}{T}\right),
\end{eqnarray}
where $\gamma$ is the Euler number and $\omega_c$ a cutoff frequency. Note that the angular integral over $|\Psi^\Gamma(\bk)|^2$ leads to one since it is normalized.

\item The second and third terms have similar traces, remembering the symmetry breaking part of the Hamiltonian is written as $\delta H (\bk) = \mathbf{s}(\bk)\cdot \boldsymbol{\sigma}$:
\begin{eqnarray}
T_2&=& Tr \left[ \hat{\Delta}(\bk)  \hat{\Delta}^\dagger(\bk) \delta H (\bk) \right] \\\nonumber
&=&Tr\left[ \left(|\Psi^\Gamma(\bk)|^2 I_2 - \mathbf{q}(\bk)\cdot\boldsymbol{\sigma}^T \right) (\mathbf{s}(\bk)\cdot \boldsymbol{\sigma})\right]\\\nonumber
&=&- 2 \mathbf{q}(\bk)\cdot \mathbf{s}(\bk),
\end{eqnarray}
and
\begin{eqnarray}
T_3&=&Tr \left[   \hat{\Delta}^\dagger(\bk) \hat{\Delta}(\bk)[\delta H(-\bk)]^T\right] \\\nonumber
&=& -Tr\left[ \left( |\Psi^\Gamma(\bk)|^2 I_2 + \mathbf{q}(\bk)\cdot\boldsymbol{\sigma}\right) (\mathbf{s}(-\bk)\cdot \boldsymbol{\sigma}^*)\right]\\\nonumber
&=&-2\mathbf{q}(\bk)\cdot \mathbf{s}(-\bk).
\end{eqnarray}

Note that these terms only contribute for the case of non-unitary pairing ($\mathbf{q}(\bk)\neq 0$). We can evaluate the sum over momenta and Matsubara frequencies for these two terms together:
\begin{eqnarray}
-2 T v \sum_{\bk,n}   \mathbf{q}(\bk)\cdot \left( \mathbf{s}(\bk)G_0+  \mathbf{s}(-\bk)\bar{G}_0\right) G_0\bar{G}_0,
\end{eqnarray}
using the fact that $G_0(\bk,i\omega_n)$ and $ \mathbf{q}(\bk)$ are even in $\bk$, we can take all the arguments in the sum over $\bk$ in the second term to be negative, and then change the sum from positive to negative $\bk$ and rewrite both terms together:
\begin{eqnarray}
-2 Tv   \sum_{\bk,n} G_0\bar{G}_0\left(G_0 +  \bar{G}_0\right) \mathbf{q}(\bk)\cdot \mathbf{s}(\bk).
\end{eqnarray}

Writing the GF explicitly:
\begin{eqnarray}
2T v \sum_{\bk,n} \frac{2 \xi(\bk)}{\left(\omega_n^2 + \xi^2(\bk)\right)}\mathbf{q}(\bk)\cdot \mathbf{s}(\bk),
\end{eqnarray}
so when performing the sum over $\bk$ as an integral over energy with a constant DOS over a symmetric interval around the Fermi energy, we find zero since the integrand is odd in energy.

\item The trace present in the fourth term can be rewritten in terms of $F(\bk)$:
\begin{eqnarray}
\delta H (\bk) \hat{\Delta}(\bk) - \hat{\Delta}(\bk) \delta H^* (-\bk) = F(\bk) (i\sigma_2),
\end{eqnarray}
multiplying by $(-i\sigma_2)$ on the right we find
\begin{eqnarray}
F(\bk)&=& \delta H (\bk) \hat{\Delta}(\bk)  (-i\sigma_2)  \\& & -     \hat{\Delta}(\bk) \delta H^* (-\bk) (-i\sigma_2),\nonumber
\end{eqnarray}
and its transpose conjugate:
\begin{eqnarray}
F^\dagger(\bk)&=&  (i\sigma_2)  \hat{\Delta}^\dagger(\bk) \delta H (\bk)    \\&& \nonumber-  (i\sigma_2) \delta H^* (-\bk)\hat{\Delta}^\dagger(\bk).
\end{eqnarray}

Evaluating the trace of $Tr|F(\bk)|^2$, we have:
\begin{eqnarray}
Tr|F(\bk)|^2 &=& Tr[   \hat{\Delta}^\dagger(\bk) \delta H (\bk)  \delta H (\bk) \hat{\Delta}(\bk)   ]\\\nonumber
&-& Tr[  \hat{\Delta}^\dagger(\bk) \delta H (\bk)   \hat{\Delta}(\bk) \delta H^* (-\bk)    ]\\\nonumber
&-& Tr[  \delta H^* (-\bk)\hat{\Delta}^\dagger(\bk)  \delta H (\bk)\hat{\Delta}(\bk)    ]\\\nonumber
&+& Tr[  \delta H^* (-\bk)\hat{\Delta}^\dagger(\bk)  \hat{\Delta}(\bk) \delta H^* (-\bk) ],
\end{eqnarray}
which can be regrouped as:
\begin{eqnarray}
Tr|F(\bk)|^2 &=& Tr[  \hat{\Delta}(\bk)\hat{\Delta}^\dagger(\bk) \left(\delta H (\bk)\right)^2  ]\\\nonumber
&+& Tr[\hat{\Delta}^\dagger(\bk) \hat{\Delta}(\bk)\left( \delta H^* (-\bk) \right)^2]\\\nonumber
&-& 2 Tr[  \hat{\Delta}^\dagger(\bk) \delta H (\bk)  \hat{\Delta}(\bk) \delta H^* (-\bk) ],
\end{eqnarray}
so we can identify:
\begin{eqnarray}
T_4=\frac{T_5+T_6}{2}-\frac{1}{2}Tr|F(\bk)|^2.
\end{eqnarray}

Before evaluating the sum over $\bk$ and Matsubara frequencies for the fourth term, we are going to evaluate the trace of the last two terms.

\item The fifth and sixth terms have equal traces:
\begin{eqnarray}
T_5&=& Tr \left[  \hat{\Delta}(\bk) \hat{\Delta}^\dagger(\bk) \left(\delta H (\bk)\right)^2  \right] \\\nonumber
&=&Tr\left[ \left(|\Psi^\Gamma(\bk)|^2 I_2 - \mathbf{q}(\bk)\cdot\boldsymbol{\sigma} \right) (\mathbf{s}(\bk)\cdot \boldsymbol{\sigma})^2\right]\\\nonumber
&=& 2 |\Psi^\Gamma(\bk)|^2 \mathbf{s}(\bk) \cdot \mathbf{s}(\bk),
\end{eqnarray}
and
\begin{eqnarray}
T_6&=&Tr \left[  \hat{\Delta}^\dagger(\bk) \hat{\Delta}(\bk)[\left(\delta H(-\bk)\right)^2]^T\right] \\\nonumber
&=&Tr\left[ \left(|\Psi^\Gamma(\bk)|^2 I_2 + \mathbf{q}(\bk)\cdot\boldsymbol{\sigma} \right) (\mathbf{s}(\bk)\cdot \boldsymbol{\sigma})^2\right]\\\nonumber
&=& 2 |\Psi^\Gamma(\bk)|^2 \mathbf{s}(\bk) \cdot \mathbf{s}(\bk),
\end{eqnarray}

Note that the same term $ 2 |\Psi^\Gamma(\bk)|^2 \mathbf{s}(\bk) \cdot \mathbf{s}(\bk)$ appears in the forth, fifth and sixth terms of the linearized gap equation, so we are going to perform the sum over momenta and Matsubara frequency of these terms alltogether:
\begin{eqnarray}\label{ZeroContrib}
-2  T v \sum_{\bk ,n} \left[ \left(G_0\bar{G}_0\right)^2+\left(G_0\right)^3\bar{G}_0\right. \\ \nonumber \left.+G_0\left(\bar{G}_0\right)^3 \right]  |\Psi^\Gamma(\bk)|^2 \mathbf{s}(\bk) \cdot \mathbf{s}(\bk).
\end{eqnarray}

The sum over momenta can be performed as an integral over energy for a constant DOS $N(0)$ plus angular integral, leading to:
\begin{eqnarray}
&&-2 T v N(0) \sum_n \int \frac{d\Omega}{4\pi} \int_{-\omega_c}^{\omega_c} d\epsilon |\Psi^\Gamma(\bk)|^2\\ \nonumber&&\times \mathbf{s}(\bk) \cdot \mathbf{s}(\bk) \frac{(i\omega_n)^2 + 3 \epsilon^2}{(-(i\omega_n)^2 + \epsilon^2)^3}\nonumber\\
&& = 4 T v N(0) \omega_c \int \frac{d\Omega}{4\pi}  |\Psi^\Gamma(\bk)|^2 \\ \nonumber &&\times\mathbf{s}(\bk) \cdot \mathbf{s}(\bk)\sum_n \frac{1}{\omega_n^3} \left[ \frac{\omega_c/\omega_n}{(1+(\omega_n/\omega_c)^2)^2}\right].
\end{eqnarray}

The term in brakets rapidly vanishes in the limit of large $\omega_c/\omega_n$, so these terms do not contribute to the gap equation in the low temperature limit.

What is left to be evaluated is the contribution of the part proportional to $Tr|F(\bk)|^2$ from the fourth term:
\begin{eqnarray}
\frac{T v}{2}  \sum_{\bk,n} Tr[|F(\bk)|^2] \left(G_0\bar{G}_0\right)^2.
\end{eqnarray}

Again, the sum over momentum can be evaluated as an integral over energy with a constant DOS and an angular integral over the FS. The integral over energy leads to:
\begin{eqnarray}
&&\frac{T v N(0)}{2}  \sum_{n}  \int \frac{d\Omega}{4\pi}  Tr|F(\bk)|^2 \\ \nonumber&&\times  \int_{-\omega_c}^{\omega_c} d\epsilon \frac{1}{\left(\omega_n^2 + \epsilon^2\right)^2}\\ \nonumber
&&= \frac{T v N(0)}{2}  \int \frac{d\Omega}{4\pi}  Tr|F(\bk)|^2 \\ \nonumber&&\times 
\sum_n  \frac{1}{\omega_n^3} \left[ \frac{\omega_c/\omega_n}{(1+(\omega_c/\omega_n)^2)} + \arctan(\omega_n/\omega_n)\right].
\end{eqnarray}

Now the term in brakets tends to the constant value $\pi/2$ in the limit of large $\omega_c/\omega_n$, and we can evaluate the sum over $n$ identifying it with the Riemann zeta function $\xi(n)$:
\begin{eqnarray}
&& \frac{\pi T v N(0)}{4}   \int \frac{d\Omega}{4\pi}  Tr|F(\bk)|^2 \sum_{n}  \frac{1}{\omega_n^3} \\ \nonumber
&&=  \frac{v N(0) }{4 (\pi T)^2}  \langle Tr|F(\bk)|^2  \rangle_{FS}\sum_{n} \frac{1}{(2n+1)^3} \\\nonumber 
&&=  \frac{7 \xi(3)}{32 \pi^2} v  N(0)  \frac{ \langle Tr|F(\bk)|^2  \rangle_{FS}}{T^2}\nonumber.
\end{eqnarray}
where we introduced the notation for the average over the Fermi surface:
\begin{eqnarray}
 \int \frac{d\Omega}{4\pi}  Tr|F(\bk)|^2  = \langle Tr|F(\bk)|^2  \rangle_{FS}.
\end{eqnarray}
\end{itemize}

In conclusion, the non-zero contributions to the gap equation come from the terms in $T_1$ and $T_4$ (only the part which carries $Tr|F(\bk )|^2$). In the absence of symmetry breaking fields it reduces to usual BCS result:
\begin{eqnarray}
1 &=&  -2 N(0) v   ln \left(\frac{2 e^\gamma}{\pi}\frac{ \omega_c}{ T_c^0}\right),
\end{eqnarray}
from which we can define the critical temperature
\begin{eqnarray}
T_c^0 = \frac{2 e^\gamma}{\pi}\omega_c e^{-1/\lambda}.
\end{eqnarray}
where $\lambda= 2 N(0) |v|$.

For the perturbed system we find:
\begin{eqnarray}\label{A29}
1 &=&  -2 N(0) v  ln \left(\frac{2 e^\gamma}{\pi}\frac{ \omega_c}{ T_c}\right)\\ &+& \frac{7 \xi(3)}{32  \pi^2} N(0) v  \frac{\left\langle Tr|F(\bk )|^2  \right\rangle_{FS }}{T_c^2},\nonumber
\end{eqnarray}
and using the result above for the original critical temperature, we can write:
\begin{eqnarray}
log \left(\frac{T_c}{T_c^0}\right)&=& -\frac{7 \xi(3)}{64 \pi^2} \frac{\left\langle  Tr|F(\bk )|^2  \right\rangle_{FS}}{(T_c^0)^2}
\end{eqnarray}
so for small perturbations:
\begin{eqnarray}
T_c&\sim& T_c^0\left(1 -\frac{7 \xi(3)}{64 \pi^2}  \frac{\left\langle  Tr|F(\bk )|^2  \right\rangle_{FS}}{(T_c^0)^2}\right),\nonumber\\
\end{eqnarray}
which is Eq.~\ref{TC} in the main text.

%%%%%%%%%%%%%%%%%%%%%%%%%%%%%%%%%%%%%%%%%
\section{Derivation of the relation of $F(\bk)$ with the suppression of $T_c$ for the multi-orbital case}\label{AppTCMO}

Following the discussion in Appendix \ref{AppTC}, here we analyze the traces that appear in the linearized gap equation up to second order in the perturbation $\delta H(\bk)$ for the multi-orbital case. Starting with the full form of the linearized gap equation, as introduced in the main text in Eq.~\ref{GapMO}:
\begin{eqnarray}
\phi^{a} &=& - T\sum_{\bk,n, \{s\}} \sum_{b,c} v^{ab}   \phi^{b*}\frac{|\phi^c|}{|\phi^b|}  \hat{\Delta}^{b\dagger}_{s_{1}s_{2}}(\bk) \\ \nonumber &&\times \, G^{bc}_{s_{2}s_{3}}(\bk,i\omega_n) \hat{\Delta}^{c}_{s_{3}s_{4}}(\bk)G^{cbT}_{s_{4}s_{1}}(-\bk,-i\omega_n).
 \end{eqnarray}

We again expand the GF, keeping only terms up to second order in $\delta H(\bk)$ in the linearized gap equation to find:
\begin{eqnarray}
\phi^{a}  &=& - T \sum_{\bk,n} \sum_{b,c, i} v^{ab} \phi^{b*} \left\{ 
G^b_{0}  \bar{G}^{b}_{0} \mathcal{T}_{1b} \right.  \\ \nonumber
&+&  G^b_{0}  \bar{G}^{b}_{0}(G^b_{0}  \mathcal{T}_{2b}+ \bar{G}^{b}_{0} \mathcal{T}_{3b})+ \frac{|\phi^c|}{|\phi^b|} \left(G^b_{0}  \bar{G}^{b}_{0} \right)^2 \mathcal{T}_{4bc}\\ \nonumber
&+&\left.    \sum_d  G^b_{0} \bar{G}^b_{0} \left(G^b_{0}  G_0^d \mathcal{T}_{5bd}
+\bar{G}^b_{0} \bar{G}_0^d  \mathcal{T}_{6bd}
\right)  \right\}.
\end{eqnarray}

Writing each trace explicitly, omitting the momentum dependence and introducing the short notation $\delta H=\delta H(\bk)$ and $\delta \bar{H}=\delta H(-\bk)$, we have:
\begin{eqnarray}
\mathcal{T}_{1b}&=& Tr \left[  \hat{\Delta}^{b\dagger}\hat{\Delta}^b \right] ,\\\nonumber
\mathcal{T}_{2b} &=& Tr \left[  \hat{\Delta}^{b\dagger} \delta H^{bb} \hat{\Delta}^{b}\right],\\\nonumber
\mathcal{T}_{3b}&=&Tr \left[  \hat{\Delta}^{b\dagger}  \hat{\Delta}^b  \delta \bar{H}^{bbT}   \right],\\\nonumber
\mathcal{T}_{4bc}&=& Tr \left[  \hat{\Delta}^{b\dagger}  \delta H^{bc} \hat{\Delta}^{c} \delta \bar{H}^{cbT}   \right],\\\nonumber
\mathcal{T}_{5bc}&=&  Tr \left[  \hat{\Delta}^b   \hat{\Delta}^{b\dagger}  \delta H^{bc}  \delta H^{cb}\right],\\\nonumber
\mathcal{T}_{6bc}&=& Tr \left[  \hat{\Delta}^{b\dagger}   \hat{\Delta}^b  \delta \bar{H}^{cbT}  \delta \bar{H}^{bcT} \right].
\end{eqnarray}

Now we analyze each term separately.

\begin{itemize}

\item The first trace is again the simplest: 
\begin{eqnarray}
\mathcal{T}_{1b} &=&2|\Psi_b^\Gamma(\bk)|^2,
\end{eqnarray}
and the term containing this trace falls back into the single-band calculation for the sum over momentum and Matsubara frequency, contributing with BCS-like terms to the gap equation.

\item Now we analyze the second and third traces. The traces can be rewritten as:
\begin{eqnarray}
\mathcal{T}_{2b}&=& -2  \mathbf{q}_b \cdot \mathbf{s}^{bb} .\\
\mathcal{T}_{3b}&=& -2  \mathbf{q}_b \cdot \bar{\mathbf{s}}^{bb}\nonumber
 \end{eqnarray}
where $ \mathbf{q}_b= i  \left( \Psi^\Gamma_b\right)^*\times  \Psi_b^\Gamma$ and $\mathbf{s}^{bb}$ refers to a diagonal block in $\delta H (\bk)$, concerning perturbations that have only intra-band effects. Again the bands decouple and we fall in the discussion for the single band case. Based on the same arguments as in Appendix \ref{AppTC}, we can combine the second and third terms and conclude that these do not contribute to the gap equation.

\item The fourth term can be written in terms of the quantity $F(\bk)$ introduced in the criterium in the main text:
\begin{eqnarray}
\mathcal{T}_{4bc}&=& \frac{\mathcal{T}_{5bc}+\mathcal{T}_{6bc}}{2}-\frac{1}{2} Tr\left[F_{bc}^\dagger  F_{bc}\right].
\end{eqnarray}

\item In the same fashion, the fifth and sixth terms are equal and can be rewritten as:
\begin{eqnarray}
\mathcal{T}_{5bc}=\mathcal{T}_{6bc}&=&2 |\Psi_b^\Gamma(\bk)|^2  \mathbf{s}^{bc}\cdot \mathbf{s}^{cb}\\ \nonumber&-&2i \mathbf{q}_b \cdot \mathbf{s}_{bc}\times \mathbf{s}_{cb}.
\end{eqnarray}

We note again that the first term in $\mathcal{T}_{4bc}$ is equal to $\mathcal{T}_{5bc}$ and $\mathcal{T}_{6bc}$ and propose to perform the sums over momentum and Matsubara frequencies of these together. Under the condition that the gaps in different bands are not very different and that the bands have a similar DOS, these terms fall into the single band discussion and do not contribute significantly to the gap equation. Note that for systems in which these considerations do not apply this term should be reconsidered.

\end{itemize}

Within these considerations, writing all terms that are left in the gap equation explicitly:
\begin{eqnarray}
\phi_a  &=& - 2T  \sum_{\bk,n}  \sum_{b} v^{ab} \phi^{b*}  |\Psi_b^\Gamma(\bk)|^2 G^b_{0}  \bar{G}_0^b\\ \nonumber
&+& \frac{T}{2}  \sum_{\bk,n}  \sum_{b,c}  \phi^{b*} \frac{|\phi^c|}{|\phi^b|}  Tr\left[F_{bc}^\dagger  F_{bc}\right] \left(G^b_{0}  \bar{G}^{b}_{0} \right)^2.
\end{eqnarray}

Using the results for the sums over momentum and Matsubara frequency from the previous Appendix, we can write the multi-band version of Eq.~\ref{A29} as:
\begin{eqnarray}\label{GapNew}
\phi^{a} &=&  -2 \sum_{b} v^{ab}  N_b(0)   \phi^{b*}  ln \left(\frac{2 e^\gamma}{\pi}\frac{ \omega_c}{T_c}\right) \\ \nonumber &+&   \frac{7 \xi(3)}{32 \pi^2} \sum_{b,c} v^{ab}N_b (0)  \phi^{b*} \frac{|\phi^c|}{|\phi^b|} \frac{ \left\langle Tr[F_{bc}^\dagger F_{bc}] \right\rangle_{FS}}{T_c^2}.
\end{eqnarray}

%%%%%%%%%%%%%%%%%%%%%%%%%%%%%%%%%%%%%%%%%
\section{SU(3) and the Gell-Mann Matrices}\label{GMM}

The Gell-Mann (GM) matrices is a set of eight $3$x$3$ traceless Hermitian matrices which are the generators of SU(3) in the fundamental representation. They are usually defined as:
\begin{eqnarray}
\lambda_1 &=&\begin{pmatrix}
0 & 1 &  0\\
1 & 0 &  0\\
0 & 0 &  0
\end{pmatrix}, \hspace{0.5cm} 
\lambda_2 =\begin{pmatrix}
0 & -i &  0\\
i & 0 &  0\\
0 & 0 &  0
\end{pmatrix},\\ \nonumber
\lambda_3 &=&\begin{pmatrix}
1 & 0 &  0\\
0 & -1 &  0\\
0 & 0 &  0
\end{pmatrix}, \hspace{0.5cm} 
\lambda_4 =\begin{pmatrix}
0 & 0 &  1\\
0 & 0 &  0\\
1 & 0 &  0
\end{pmatrix},\\ \nonumber
\lambda_5 &=&\begin{pmatrix}
0 & 0 &  -i\\
0 & 0 &  0\\
i & 0 &  0
\end{pmatrix}, \hspace{0.5cm} 
\lambda_6 =\begin{pmatrix}
0 & 0 &  0\\
0 & 0 &  1\\
0 & 1 &  0
\end{pmatrix},\\ \nonumber
\lambda_7 &=&\begin{pmatrix}
0 & 0 &  0\\
0 & 0 &  -i\\
0 & i &  0
\end{pmatrix}, \hspace{0.5cm} 
\lambda_8 =\frac{1}{\sqrt{3}}\begin{pmatrix}
1 & 0 &  0\\
0 & 1 &  0\\
0 & 0 & -2
\end{pmatrix},
\end{eqnarray}
and follow the commutation relations:
\begin{eqnarray}
[\lambda_a, \lambda_b]=2i f_{abc} \lambda_c,
\end{eqnarray}
where $f_{ijk}$ are completely antisymmetric tensors.

Some useful properties of these matrices are:
\begin{eqnarray}
Tr[\lambda_a]&=&0,\\
Tr[\lambda_a\lambda_b]&=& 2\delta_{ab},\\
Tr[\lambda_a\lambda_b\lambda_c]&=& 2h_{abc},\\
Tr[\lambda_a\lambda_b\lambda_c\lambda_d]&=& \frac{\delta_{ab}\delta_{cd}}{3} + h_{abi} h_{icd},
\end{eqnarray}
where $h_{abc}=d_{abc}+if_{abc}$, with $d_{abc}$ a fully symmetric tensor. The fully symmetric and antisymmetric tensors can be written in terms of the GM matrices defined above as:
\begin{eqnarray}
d_{abc}=\frac{1}{4}Tr[\lambda_a\{\lambda_b,\lambda_c\}],\\
f_{abc}=-\frac{i}{4}Tr[\lambda_a[\lambda_b,\lambda_c]].
\end{eqnarray}

%%%%%%%%%%%%%%%%%%%%%%%%%%%%%%%%%%%%%%%%%%%

\end{document}